# Photonic crystal waveguide crossing based on transformation optics

M. M. Gilarlue[a], S. Hadi Badri[a],*

[a] Department of Electrical Engineering, Sarab Branch, Islamic Azad University, Sarab, Iran

* sh.badri@iaut.ac.ir

## Abstract

The absolute instruments like the Maxwell's fisheye (MFE) lens with aberration-free imaging properties have found interesting applications such as waveguide crossing. The flat wavefront of an optical wave in the waveguide does not match with the circular wavefront of the circular MFE lens at its edge, hence, we design and study the performance of a square MFE lens as photonic crystal waveguide crossing medium. We also have truncated the square MFE lens to a cross-shaped lens to squeeze it inside the crossing waveguides, therefore, practically no extra footprint is consumed by the truncated MFE lens. The numerical simulations show that graded photonic crystal-based implementation of the truncated MFE lens provides a bandwidth of 186 nm covering the entire S- and C-bands and partially covering the E- and L-bands of optical communication. The crosstalk levels are lower than -18 dB while the average insertion loss is 0.32 dB in the C-band.

## 1. Introduction

Planar waveguide intersections with high bandwidth, low crosstalk, and low insertion loss are key elements in guiding complex lightwave traffic through highly dense photonic integrated circuits (PICs) without a need for vertical structure similar to the electronic integrated circuits. The ability of the photonic crystals (PhCs) in controlling the optical waves has made PhC structures an indispensable part of PICs. Resonant cavities located in the middle of the crossing waveguides to reduce the crosstalk have been studied [1, 2]. Waveguide crossings based on coupled cavity waveguides [3, 4], Wannier basis design [5], topology optimization [6], and self-collimation phenomenon [7] have also been proposed. Simultaneous transmission of optical signals with the same wavelength through the waveguide crossings designed by resonant cavity is impossible since they interfere with each other [3]. Moreover, due to the trade-off between transmission and bandwidth in the structures that rely on resonant cavities, waveguide crossings based on cavities offer limited bandwidth. Recently, imaging properties of the Maxwell's fisheye (MFE) lens has been utilized to design waveguide crossing media allowing the intersection of multiple waveguides [8, 9]. The MFE lens creates an image of the point source on its surface on the diametrically opposite side of the lens.

According to the Fermat's principle, light rays generally follow the shortest optical path, which is the product of the refractive index and physical path. Therefore, in gradient index (GRIN) medium the light rays follow the curved path and it can bend light in a desired direction. Novel applications have been proposed for GRIN lenses such as Luneburg [10, 11], MFE [12, 13], and Eaton [14] lenses. The mathematical technique to design GRIN media is transformation optics (TO) [15, 16] and it has been used to design various devices [17, 18]. Form-invariance of Maxwell's equations is the basis for TO [19, 20]. The correspondence between material properties and coordinate transformation had been known since 1961 [21], but it was not appreciated since the required material properties were not feasible. Advances in metamaterials introduced new possibilities to implement synthetic structures with permittivities and permeabilities not available in nature, therefore form-invariance of Maxwell's equations has been brought back to the spotlight [22-24] and consequently TO has been developed [25]. Any optical device with the given geometry, virtual domain, can be transformed into infinite number of new geometries, physical domain, with the same optical response. As the deformation of the physical domain increases, the material properties needed for its realization becomes more extreme. In general, artificially structured metamaterials incorporating resonant elements have significant absorption losses and limited bandwidth due to the large frequency dispersion. Therefore, certain restrictions have been applied to quasi conformal transformation optics (QCTO) in order to design broadband components requiring only non-resonant, isotropic, non-magnetic, and all-dielectric metamaterials [26, 27].

In this paper, we design a square Maxwell's fisheye lens as a waveguide intersection medium based on QCTO. Two designs are presented for PhC waveguide crossings. In the first design, the parallel waveguides enter the waveguide intersection with bandwidth of 82 *nm* covering 1510-1592 *nm* while the average insertion loss is 0.33 dB and crosstalk levels are lower than -24 dB in the C-band. In the second design, the waveguides intersect each other vertically where the truncated square MFE lens is squeezed inside the crossing. In this case, the crosstalk levels are lower than -15 dB in the wide wavelength range of 1422-1608 *nm* while the average insertion loss in the C-band is 0.32 dB.

## 2. Simple intersection of PhC waveguides

The photonic crystal with a square lattice of silicon (Si) rods in an air background with the lattice constant of $a$=540 nm and radius of $r_{rod}$=0.185$a$ is considered [28]. The band structure diagram of this PhC structure, as shown in Fig. 2(a), indicates that it has two photonic bandgaps (PBGs) in transverse magnetic (TM) mode where the rods are parallel to the electric field. The first PBG is at about $0.3 < a/\lambda < 0.44$ corresponding to $1227 nm < \lambda < 1800 nm$ covering the entire optical communication bands in TM mode. Simple intersection of two PhC waveguides suffers from high insertion and return losses as well as high levels of crosstalk. As displayed in Fig. 1(b), the crosstalk levels are -6.0 dB while the return loss is 4.5 dB and insertion loss is 5.3 dB at the wavelength of 1550 nm.

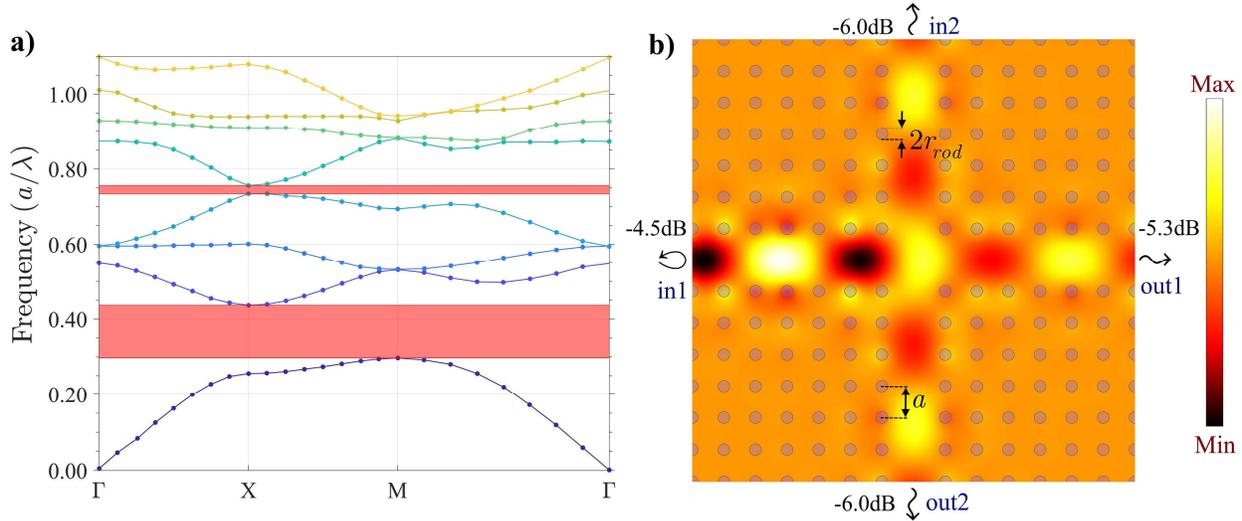

Fig. 1. a) The bandgap structure of the considered PhC structure in the TM mode. b) The electric field distribution of the TM mode light injected from *in1* in the simple intersection of PhC waveguides at the wavelength of 1550 nm. The crosstalk levels and return loss are considerably high in the simple intersection of PhC waveguides.

## 3. Quasi conformal transformation optics

In this paper, a circular MFE lens as the virtual domain is transformed to a square lens in the physical domain [29, 30]. The QCTO is based on boundary transformation instead of point to point mapping. Outer boundaries of the virtual domain are mapped to the outer boundaries of the physical domain. Boundary-based mapping could be conformal or non-conformal [31]. The QCTO is performed in two steps. The virtual domain is mapped to a rectangle intermediate domain in the first step. Dimensions and aspect ratio of the intermediate domain are determined by the Dirichlet boundary conditions of the virtual domain. Then the physical domain is mapped to an another intermediate domain in the second step. Then the mapping in the second step is reversed to transform the intermediate domain to the physical domain. To perform the QC mapping from virtual to physical domain, these steps should be followed sequentially. These two intermediate domains should have the same area and conformal module which is the ratio of the lengths of the two adjacent sides of a domain [32, 33]. The transformed orthogonal coordinate grids of the virtual and physical domains are shown in Fig. 2(a) and (b), respectively. Four corner-like protrusions are added to the circular virtual domain to facilitate the orthogonal grid generation [34, 35]. It should be noted that since in our design the intermediate and physical domains are both rectangles, the first step of mapping gives us the desired transformation mapping. The transformation from virtual domain $(x', y', z')$ to physical domain $(x, y, z)$ is described with

$$x = x(x', y') \;,\;\; y = y(x', y') \;,\;\; z=z' \tag{1}$$

which is mapped to the material properties by

$$\varepsilon = \frac{\boldsymbol{J}\varepsilon'\boldsymbol{J}^T}{|\boldsymbol{J}|} \;,\;\; \mu = \frac{\boldsymbol{J}\mu'\boldsymbol{J}^T}{|\boldsymbol{J}|} \tag{2}$$

where $\varepsilon$ and $\mu$ are transformed permittivity and permeability in the physical domain, $\boldsymbol{J}$ is the Jacobian transformation matrix between the virtual and physical domains, and the spaitial distribution of permittivity ($\varepsilon'$) and permeability ($\mu'$) in the virtual domain is

$$\mu' = 1 \;,\;\; n_{vir}(r') = \sqrt{\varepsilon'} = \begin{cases} n_{lens}(r') = \dfrac{2 \times n_{edge}}{1+(r'/R_{lens})^2} & r' \le R_{lens} \\ n_{edge} & r' > R_{lens} \end{cases} \tag{3}$$

where $R_{lens}$=2.46 µm is the radius of the lens, $r'$ is the radial distance from the center of the lens, and $n_{edge}$ is the refractive index of the lens at its edge and is chosen as unity. The side length of the square is 4.05 µm. In QCTO, the equation (3) is reduced to below equation by Cauchy-Riemann condition:

$$\mu = 1 \;,\;\; \varepsilon = \frac{n_{vir}^2(r)}{|\det \boldsymbol{J}|} \tag{4}$$

The refractive index of the virtual and physical domains are shown in Fig. 2(c) and (d), respectively.

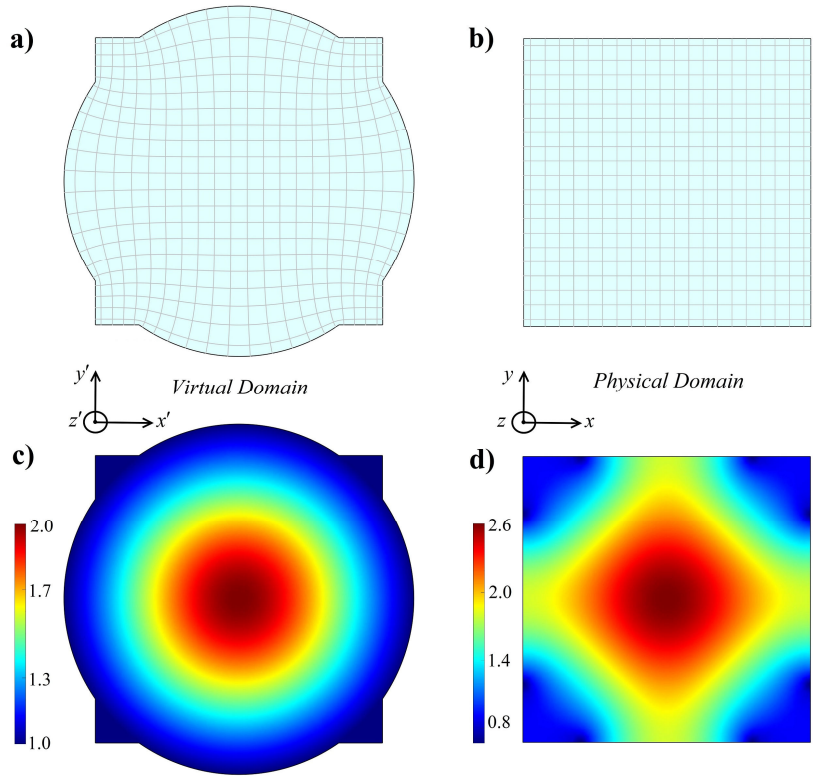

Fig. 2. The orthogonal coordinate grids of a) the virtual and b) the physical domains. The refractive index profile of c) the virtual and d) the physical domains.

## 4. Implementation of the MFE lens

In non-resonant metamaterials, the size of the structure is considerably smaller than the operating wavelength. Therefore, the Mie resonance and Bragg diffraction are negligible in these metamaterials. In our design, the effective medium theory is utilized to implement a medium with a range of refractive-indices between the refractive indices of the constituent materials, i.e., Si and air. Within the long-wavelength limit, the subwavelength structure can be regarded as a homogeneous medium. Graded photonic crystal (GPC) is an example of effective medium. GPC can be used to implement a GRIN medium by changing the photonic crystal's constitutive parameters such as the refractive index of the rods, the geometry of the unit cell, or the radii of rods [36, 37]. The advantage of devices based on GPC is that these structures are relatively broadband, since GPC structures can be dispersionless for a wide range of wavelengths [38]. Various optical devices have been implemented by GPC [39, 40]. The effective refractive index of the GPC structure depends on the electric field direction with respect to the arrangement of the inclusion rods in the host material. In order to implement the MFE lens with GPC, the lens is divided into cells with appropriate size and shape. Then a cylindrical rod is placed at the center of each cell. For the TM mode, the effective refractive index is approximated by [37]

$$\varepsilon_{eff,TM} = f_{rod}\varepsilon_{rod} + (1-f_{rod})\varepsilon_{host} \tag{5}$$

where $\varepsilon_{host}$, $\varepsilon_{rod}$, and $\varepsilon_{eff,TM}$ are the permittivities of the host, rod, and effective medium for TM mode, respectively. The filling factor, $f_{rod}$, is the volume fraction occupied by the rods. The effective refractive index of the GPC structure is controlled by varying the filling factor of the rods. The above equation can be rearranged as

$$f_{inc} = \frac{\varepsilon_{eff,TM} - \varepsilon_{host}}{\varepsilon_{rod} - \varepsilon_{host}} \tag{6}$$

The filling factor for the ij-th cell is $f_{inc} = A_{rod}/A_{ij}$ where $A_{rod} = \pi r_{rod,ij}^2$ and $A_{ij} = a_{GPC}^2$ is the area of the ij-th square cell. $a_{GPC}$ is the lattice constant of the GPC structure. The $r_{rod,ij}$ is the radius of the rod placed at the center of the ij-th cell and is calculated by

$$r_{rod,ij} = a_{GPC}\sqrt{\frac{(\varepsilon_{eff,ij}^{TM} - \varepsilon_{host})}{\pi(\varepsilon_{rod} - \varepsilon_{host})}} \tag{7}$$

Our approximation method from three-dimensional (3D) model to two-dimensional (2D) model is based on simple area averaging [41].

## 5. Numerical simulation and discussion

The 2D simulations are performed by Comsol Multiphysics™ to evaluate the performance of the designed waveguide crossings. The Comsol's built-in model of Si is used in simulations. In the following subsections, the design and performance of the square and cross-shaped truncated MFE lenses are discussed. The computation domain of simulations is similar to our previous works [8, 9], so it is not presented graphically in this paper. The rectangular ports are used to calculate the scattering parameters of the waveguide crossings. PhC-based perfectly matched layers (PMLs) are placed behind the ports to truncate the computation domain and reduce spurious reflection from the ports. In the PhC-based PML, the PML layer is placed behind the port and it is surrounded by PhC.

## 5.1 Square MFE lens

As displayed in Fig. 3(a), the square MFE lens of Fig. 2(d) is implemented by GPC and used as crossing medium of two PhC waveguides. The distance between the edge of the GPC and main PhC structures is denoted by $d$ in Fig. 3(a). The lattice constant of the GPC, $a_{GPC}$, and $d$ are used to optimize the performance of the waveguide crossings. As shown in Fig. 3(b), the two parallel waveguides intersect the square lens orthogonally at its side edges. The electric field distribution of the optical wave with a wavelength of 1550 nm propagating from *in1* to *out1* is also illustrated in Fig. 3(b). The transmission, reflection, and crosstalk levels, at 1550 nm, are also depicted in this figure. In the corners of the square MFE lens, the refractive index is near or lower than unity. Our simulations show that replacing these values with unity has very limited effect on the performance of the intersection, so there are no rods in these corners in our implementation. In this design, the waveguide bends are incorporated in the lens i.e., there is no need to bend the waveguides entering the crossing medium. Hence, the footprint required for implementing bends before and after the intersection is saved. In previous designs [1-7], waveguides should be bent to adjust the crossing angle. The square lens of Fig. 2(d) occupies about 30% less footprint compared to the circular lens of Fig. 2(c). The footprint of the waveguide crossing based on square MFE lens is 4.32µm×4.32µm. Furthermore, in the waveguide crossing based on the square MFE lens, there is no need to relocate the rods near the edge of the lens. However, in the waveguide crossing based on circular MFE lens, some rods of the main photonic crystal lattice near the edge of the lens should be relocated [8, 9].

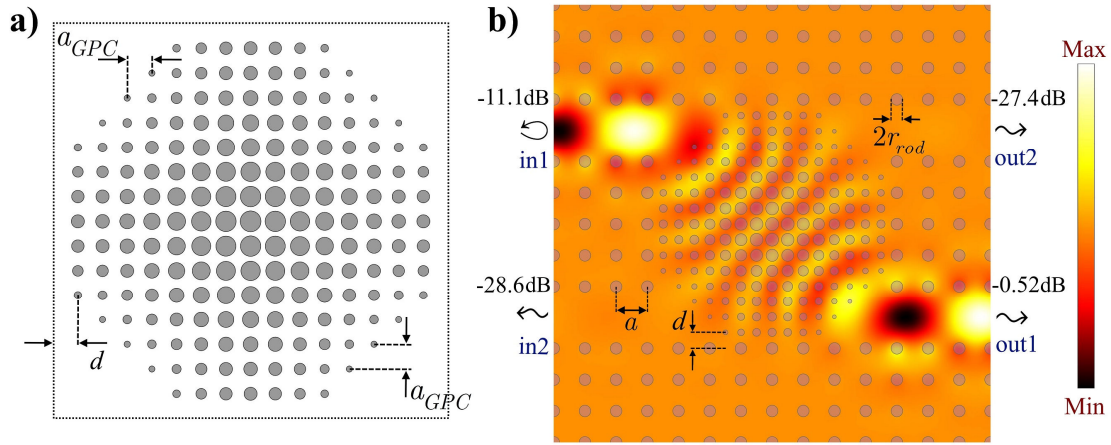

Fig. 3. a) GPC implementation of the square MFE lens displayed in Fig. 2(d). b) Electric field distribution of the TM mode wave propagating from *in1* to *out1* at 1550nm through the square MFE lens. Parameters of the main PhC lattice are *a*=540 nm and $r_{rod}$=0.185*a* while for the GPC-based lens $a_{GPC} = 269nm$ and *d*=280 nm.

We also present the transmission diagrams of three implementations of the square lens with different $a_{GPC}$ and *d* in Fig. 4. These implementations are optimized to cover the C-band by trying different values of $a_{GPC}$ and *d*. In the MFE lens with $a_{GPC} = 292nm$ and *d*=260 nm, the diameter of the largest and smallest rods are 240 and 80 nm, respectively. While the diameters of the largest and smallest rods are 225 and 67 nm, respectively, for the structure with $a_{GPC} = 269nm$ and *d*=280 nm. In the square MFE lens with the smallest grid size, $a_{GPC} = 255nm$, the distance between the GPC and main PhC is *d*=245 nm. In this structure, the largest and smallest rods are 212 and 50 nm, respectively. As the lattice constant of the GPC structure increases its performance in lower wavelengths degrades considerably. On the other hand, as $a_{GPC}$ increases the size of rods increases resulting in easier fabrication.

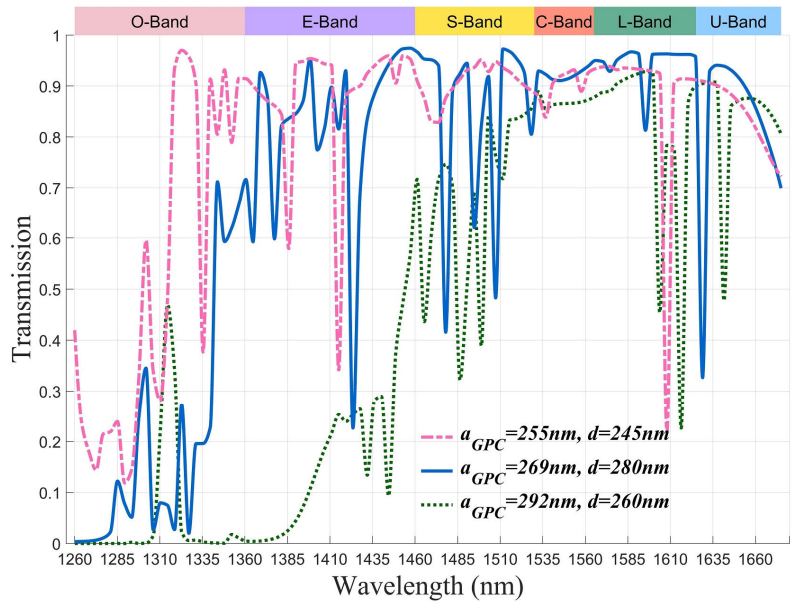

Fig. 4. Comparison between the performance of waveguide crossings based on the square MFE lens. The three implementations of the square MFE lens are optimized to cover the C-band.

The scattering parameters of the square MFE lens with $a_{GPC} = 269nm$ and *d*=280 nm are shown in Fig. 5. Four passbands with crosstalk levels below -15 *dB* are highlighted in this figure. Criteria for choosing the passbands are usually based on 3 dB transmission, however, in Fig. 5 the crosstalk levels lower than -15 dB are also considered to specify the bands with higher isolation. The first passband has a bandwidth of 43 nm covering 1431-1474 nm. The maximum return and insertion losses in this passband are 9.2 and 0.78 dB, respectively. The second passband, compared to the other passbands, has the widest bandwidth of 82 nm covering 1510-1592 nm with maximum return and insertion losses of 8.7 and 0.94 dB, respectively. The third passband has a bandwidth of 27 nm covering 1598-1625 nm with maximum insertion and return losses of 0.18 and 25 dB. In this passband the crosstalk levels are below -25 dB. The last passband covers 1633-1675 nm. The crosstalk levels are below -18 dB while the maximum return and insertion losses of 6.5 and 1.5 dB are achieved for this passband, respectively. In the C-band, the average insertion loss is 0.33 dB while the crosstalk levels are lower than -17 dB. It should be noted that the insertion loss in some parts of E- and S-bands is below half power criteria, however, the crosstalk levels are higher than -15 dB. Moreover, there are some narrow passbands which are not specified in this figure.

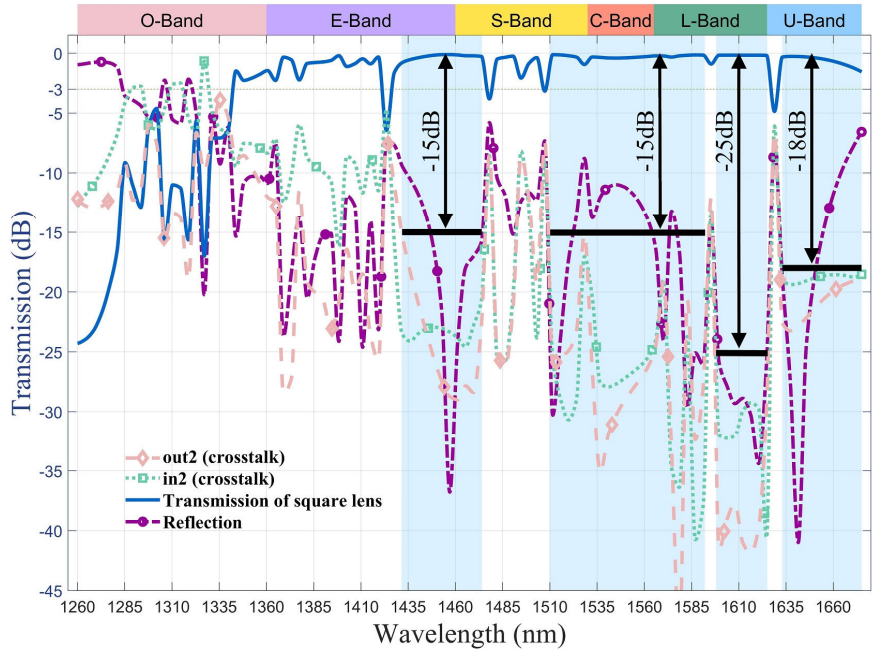

Fig. 5. Scattering parameters of the designed waveguide crossing based on the square MFE lens with $a_{GPC} = 269nm$ and $d$=280 nm. The passbands with crosstalk levels lower than -15dB are highlighted.

## 5.2 Truncated MFE lens

In this section we design a waveguide crossing for orthogonal waveguides. When the waveguides cross each other orthogonally, there is no need for the wavefront to bend 45° and go through the lens diagonally like the design of Fig. 3. So we suppose that the waveguides arrive at 45° angle to the corners of the square lens and leave it at the same direction. As shown in Fig. 6(a), we truncate the square MFE lens to a cross-shaped MFE lens. The side length of the square MFE lens is 2.46 μm in Fig. 6(a). The GPC-based implementation of the truncated lens is shown in Fig. 6(b). The size of the truncated lens is chosen to squeeze it inside the intersecting PhC waveguides so virtually the truncated lens imposes no extra footprint. The electric field distribution of the optical wave with a wavelength of 1550 nm propagating through the truncated MFE lens is shown in Fig 7. Similar to our previous design the $a_{GPC}$ and $d$ are changed to optimize the transmission in the waveguide crossing. For the design shown in Fig. 7, $a_{GPC} = 209nm$ and $d$=245 nm. The parameter

$d$ is shown in Fig. 7. The maximum and minimum diameters of the rods in the GPC structure are 165 and 47 nm, respectively.

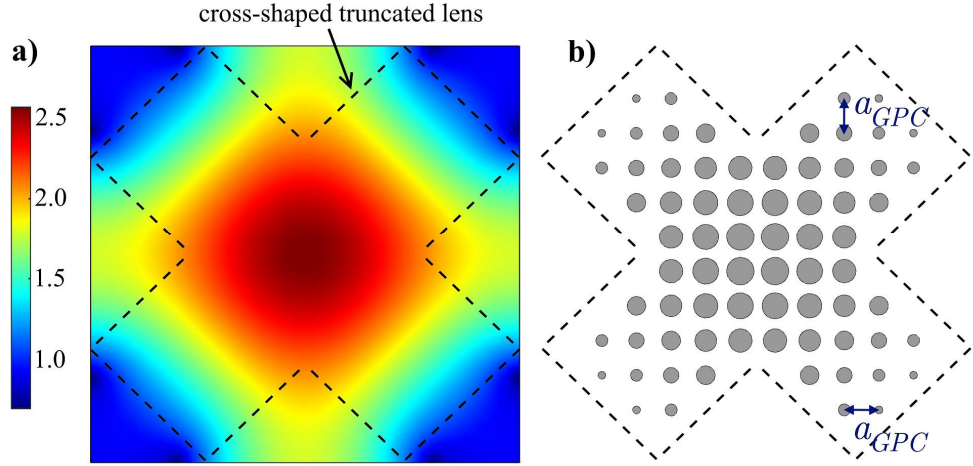

Fig. 6. a) Cross-shaped truncated MFE lens as waveguide crossing medium. b) GPC-based implementation of the truncated lens.

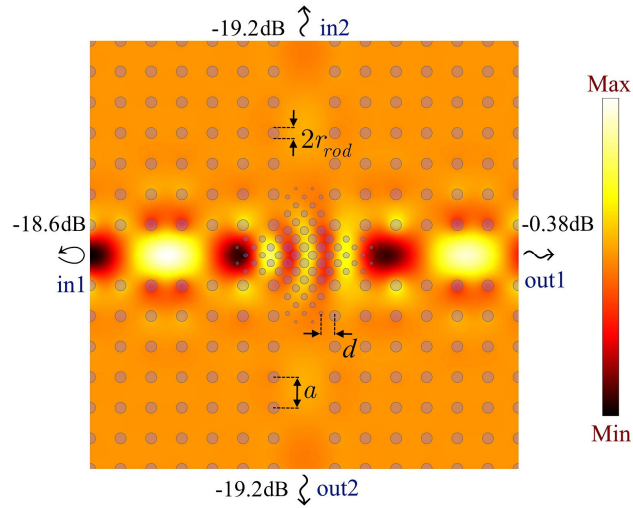

Fig. 7. The electric field distribution of the TM mode optical wave with a wavelength of 1550nm propagating through the cross-shaped truncated MFE lens. Parameters of the main PhC lattice are $a$=540 nm and $r_{rod}$=0.185$a$.

The reflection, crosstalk, and transmission of the truncated MFE lens is shown in Fig. 7. Two passbands with crosstalk levels below -15 dB are highlighted in this figure. The first passband provides a bandwidth of 64 nm partially covering the O- and E-bands with maximum return and insertion losses of 5.0 and 1.1 dB. The second passband covers from 1422 to 1607 nm with maximum return and insertion losses of 7.0 and 3 dB. In the C-band, the maximum return and insertion losses of 9.0 and 0.41 dB are achieved. The footprint of the waveguide crossing based on truncated MFE lens is reduced significantly compared to the waveguide crossing of Fig. 3. Besides the substantial reduction of the footprint, the waveguide crossing based on truncated lens has wider bandwidth compared to the square MFE lens. This stems from the fact that in square MFE lens the wavefront mismatches between the lens and entering/exiting waveguides [13] and the wavefront rotates by 45° to pass the lens. However, in the truncated lens the flat wavefront of the waveguide matches with the truncated lens.

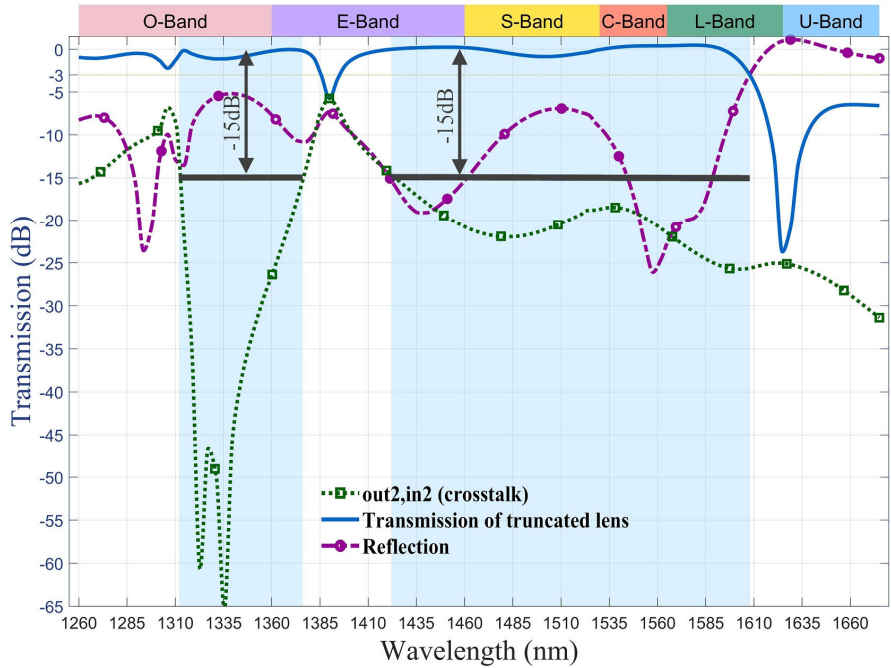

Fig. 8. Scattering parameters of the designed waveguide crossing based on the truncated MFE lens of Fig. 7. The passbands with crosstalk levels lower than -15dB are highlighted.

### 5.3 Comparison with previous studies

In order to compare our results with previous studies, we have extracted the results of references [1-7]. Fig. 7 shows the transmission of these references and the truncated MFE lens. To visualize the results of these references, the 3-dB passband and the corresponding bandwidth for each of them are illustrated in this figure. The insertion loss of reference [3] is higher than 3-dB so we have not determined a 3-dB passband for this reference. For some of these references, the passband extends beyond the optical communication bands which are not shown in Fig. 7. The truncated MFE lens has two 3-dB passbands covering 1260-1385, and 1396-1607 nm. The widest 3-dB passband of the truncated lens has 211 nm bandwidth. The only study with comparable bandwidth is reference [7], however, with larger footprint. It should be mentioned that some references use normalized frequency based on lattice constant. When the lattice constant is not given in the reference, we choose the lattice constant corresponding to the center of the photonic bandgap or waveguide crossing's passband. We also compare our design with previous studies in Table 1. Crossing mechanism, lattice type, crosstalk levels, and footprint are compared in this table. Previous designs mostly rely on resonant cavities resulting in limited bandwidth of the waveguide crossings. The bandwidth of these crossings can be increased by decreasing the Q-factor of the cavities and consequently sacrificing the transmission levels. Their footprints are given based on the lattice constant ($a$) of the PhC. Cavity resonators have also been utilized to design plasmonic waveguide crossings [42-44]. We also considered fabrication imperfections by introducing random deviation from the designed GPC-based lens. Random 10% deviation in the radii of GPC rods results in the maximum excess insertion loss of 0.6 dB.

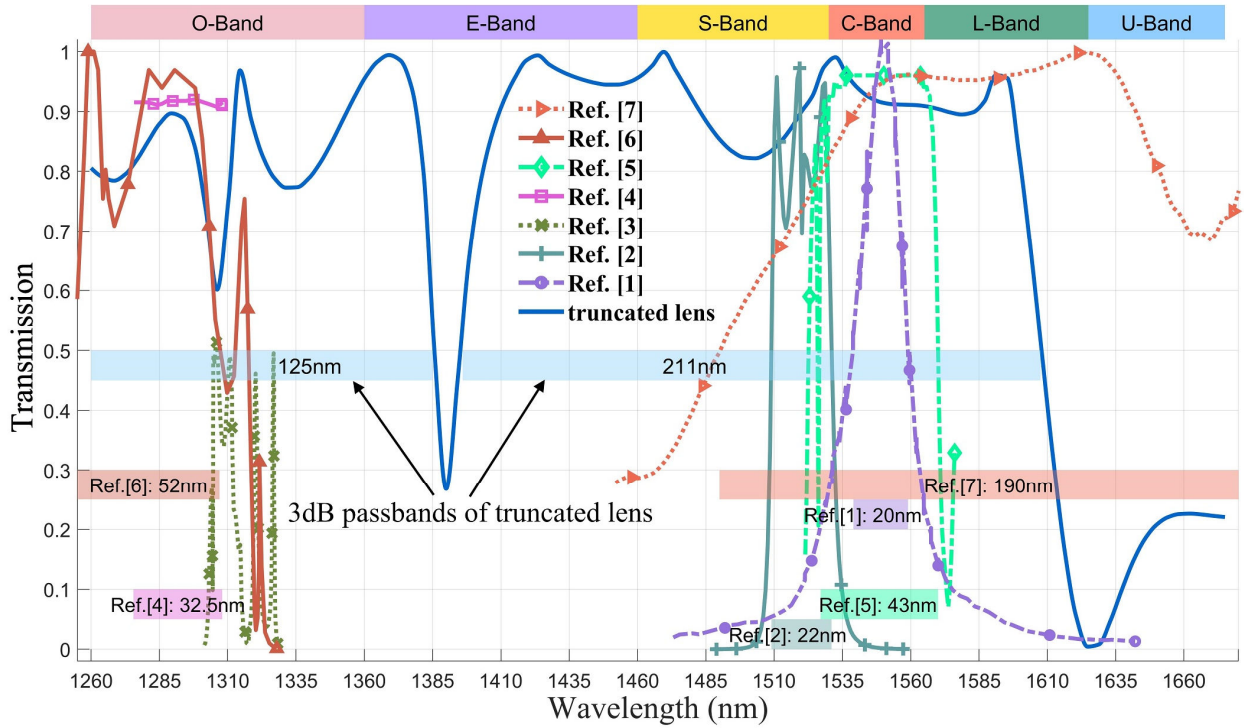

Fig. 9. Comparison of the waveguide crossing based on the cross-shaped truncated MFE lens's transmission with references [1-7]. The 3-dB passband and the corresponding bandwidth are illustrated for the designed waveguide crossing and the references.

Table 1. Comparison of the proposed designs and other waveguide crossings

| Ref. | Year | Crossing mechanism | Lattice type | Cross-talk(dB) | Footprint |
|---|---|---|---|---|---|
| [1] | 1998 | Cavity | Square | -64 | $2a \times 2a$ |
| [2] | 2009 | Cavity | Square | -46 | $8a \times 8a$ |
| [3] | 2002 | Cavity | Triangular | -42 | $7a \times 7a$ |
| [4] | 2007 | Cavity | Square | -30 | $2a \times 2a$ |
| [5] | 2005 | Cavity | Square | -40 | $8a \times 8a$ |
| [6] | 2006 | Topology optimization | Triangular | -22 | $5a \times 6a$ |
| [7] | 2019 | Self-collimation | Square | -15 | $8a \times 8a$ |
| square MFE lens | this work | MFE lens | Square | -15 | $8a \times 8a$ |
| truncated square MFE lens | this work | MFE lens | Square | -15 | $6a \times 6a$ |

# 6. Conclusion

TO is a unique method to manipulate the flow of light. The square MFE lens is designed with QCTO as crossing medium. Two waveguide crossings are designed based on the square MFE lens and numerically evaluated. In the first design, two parallel waveguides enter the intersection where the average insertion loss is 0.33 dB while the crosstalk levels are lower than -17 dB in the C-band. In the second design, the intersection angle of waveguides is 90°. In this design, the square MFE lens is squeezed inside the PhC waveguides by truncating it to a cross-shaped lens. Therefore, cross-shaped truncated MFE lens practically imposes no extra footprint. The maximum return and insertion losses of 9.0 and 0.41 dB are achieved in the C-band. The wideband waveguide crossing based on the truncated MFE lens has a 3-dB bandwidth of 211 nm.